# Role of AI in Theranostics: Towards Routine Personalized Radiopharmaceutical Therapies

Julia Brosch-Lenz[1], Fereshteh Yousefirizi[1], Katherine Zukotynski[2], Jean-Mathieu Beauregard[3,4], Vincent Gaudet[5], Babak Saboury[6,7,8], Arman Rahmim[1,9,10], Carlos Uribe[10,11,*]

[1]Department of Integrative Oncology, BC Cancer Research Institute, Vancouver, BC, Canada
[2]Department of Medicine and Radiology, McMaster University, Hamilton, ON, Canada
[3]Department of Radiology and Nuclear Medicine, and Cancer Research Centre, Université Laval, Québec City, QC, Canada
[4]Department of Medical Imaging, and Research Center (Oncology Axis), CHU de Québec - Université Laval, Québec City, QC, Canada
[5]Department of Electrical and Computer Engineering, University of Waterloo, Waterloo, ON, Canada
[6]Department of Radiology and Imaging Sciences, Clinical Center, National Institutes of Health, Bethesda, MD, USA
[7]Department of Computer Science and Electrical Engineering, University of Maryland Baltimore County, Baltimore, MD, USA
[8]Department of Radiology, Hospital of the University of Pennsylvania, Philadelphia, PA, USA
[9]Department Physics, University of British Columbia, Vancouver, BC, Canada
[10]Department of Radiology, University of British Columbia, Vancouver, BC, Canada
[11]Department of Functional Imaging, BC Cancer, Vancouver, BC, Canada
*Corresponding author: curibe@bccrc.ca

**Key points:**

1. AI has shown promising applications in quantitative imaging required for dosimetry.

2. Segmentation of organs and tumors, the most time consuming task in the dosimetry workflow, can be automated using AI.

3. Using the theranostic approach, AI models that predict absorbed dose and therapy outcomes might play a key role in personalizing RPTs.

4. AI has significant potential to improve accuracy and reduce times for routine implementation of patient-specific dosimetry in RPTs.

**Disclosure statement:** The authors have nothing to disclose.

## Abstract

We highlight emerging uses of artificial intelligence (AI) in the field of theranostics, focusing on its significant potential to enable routine and reliable personalization of radiopharmaceutical therapies (RPTs). Personalized RPTs require patient-individual dosimetry calculations accompanying therapy. Image-based dosimetry needs: 1) quantitative imaging; 2) co-registration and organ/tumor identification on serial and multimodality images; 3) determination of the time-integrated activity; and 4) absorbed dose determination. AI models that facilitate these steps are reviewed. Additionally we discuss the potential to exploit biological information from diagnostic and therapeutic molecular images to derive biomarkers for absorbed dose and outcome prediction, towards personalization of therapies. We try to motivate the nuclear medicine community to expand and align efforts into making routine and reliable personalization of RPTs a reality.

## Introduction

Radiopharmaceutical therapy (RPT) has shown promise in the treatment of various cancer types [1]. Metabolic processes or specific receptors serve as targets for the design of appropriate radiopharmaceuticals. The principle of "theranostics", in the context of nuclear medicine (Figure 1), uses pairs of radiopharmaceuticals to meet and explore both, therapeutic and diagnostic purposes (i.e. "thera-nostic"). The pharmaceuticals bind to the same target and can be radiolabeled with either a therapeutic (e.g. beta or alpha emitting) or diagnostic imaging (e.g. positron or gamma emitting) radionuclide [2]. This approach allows us to "see what we treat" and "treat what we see" at the molecular level.

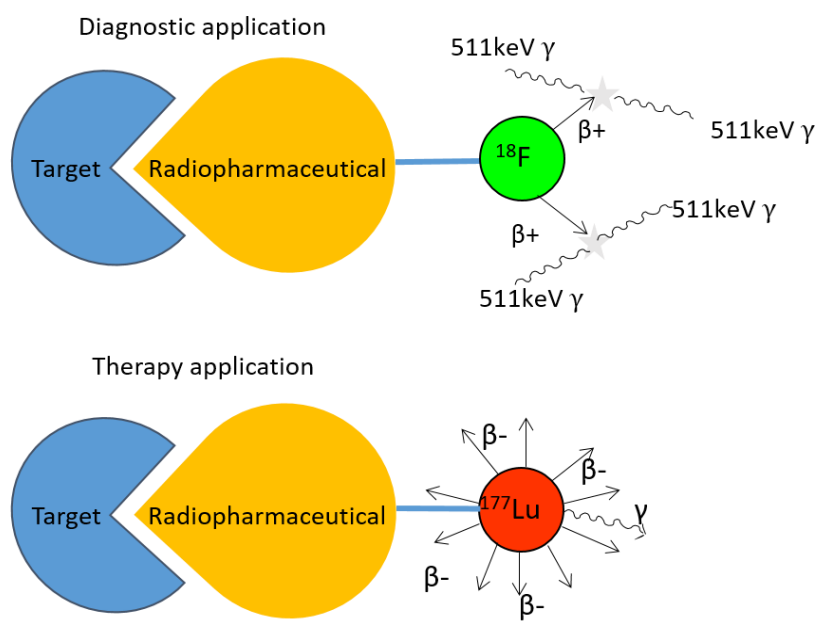

*Figure 1: An example of the principle of theranostics in nuclear medicine. Here, a radiopharmaceutical (yellow) developed to bind to the target (blue) can be labeled both with Fluorine-18 ($^{18}F$) for diagnostic imaging purposes and Lutetium-177 for therapeutic procedures.*

Two examples of recent radiopharmaceutical developments include targeting (i) the somatostatin receptors for the diagnosis and treatment of neuroendocrine tumors (NET) [3], and (ii) the prostate specific membrane antigen (PSMA) to diagnose and treat metastatic castration resistant prostate cancer (mCRPC) [4]. These are major frontiers in nuclear medicine, with significant existing and upcoming investments and efforts [5]. So far, the procedure guidelines from the International Atomic Energy Agency (IAEA), the Society of Nuclear Medicine and Molecular Imaging (SNMMI), and the European Association of Nuclear Medicine (EANM) have suggested the use of several cycles of a fixed therapeutic injection containing an activity of 7.4 GBq when Lutetium-177 ($^{177}Lu$) labelled compounds are used for NETs [6] or mCRPC treatments [7]. For NETs, [$^{177}Lu$]Lu-oxodotreotide has been approved by regulatory agencies to be used with a fixed activity of 7.4 GBq as the only option [8], and a similar framework is expected for [$^{177}Lu$]Lu-PSMA-617 for mCRPC in the near future.

Treatment planning, however, should consider individual factors such as the patient's weight and height, the tumor burden, overall patient's health condition as well as personal preferences and values. The organs at risk (OARs) tolerance to radiation and function as well as the patient-specific biological clearance and uptake of the radiopharmaceutical are further of substantial interest in personalized therapy. The key prerequisite for personalizing RPTs are routine and reliable dosimetry calculations. If dosimetry accompanies RPT, relationships between tumor and OAR radiation absorbed dose and therapy outcomes could be derived, providing evidence for adaptive treatment planning in clinical practice [9].

The present state of RPTs (Figure 2A) involves a diagnostic scan that is used by the physician to determine if a patient is suitable for therapy. If the patient expresses the target of interest, it is then referred to several cycles of therapy. Inter-therapy imaging is performed to qualitatively assess the performance of the treatment (e.g. to visualize distribution of the therapeutic radiopharmaceutical). To date, the use of routine post-therapy dosimetry has been hindered by its complexity and immense workload for physicians, technologists, and medical physicists. Thus, to be adopted into routine clinical practice, not only does the technique need to be accurate, but also practical. Any development that simplifies, automates, or accelerates the steps within the dosimetry workflow would be likely to increase implementation of personalized medicine. Artificial intelligence (AI) may be a game changer in supporting and facilitating the dosimetry workflow.

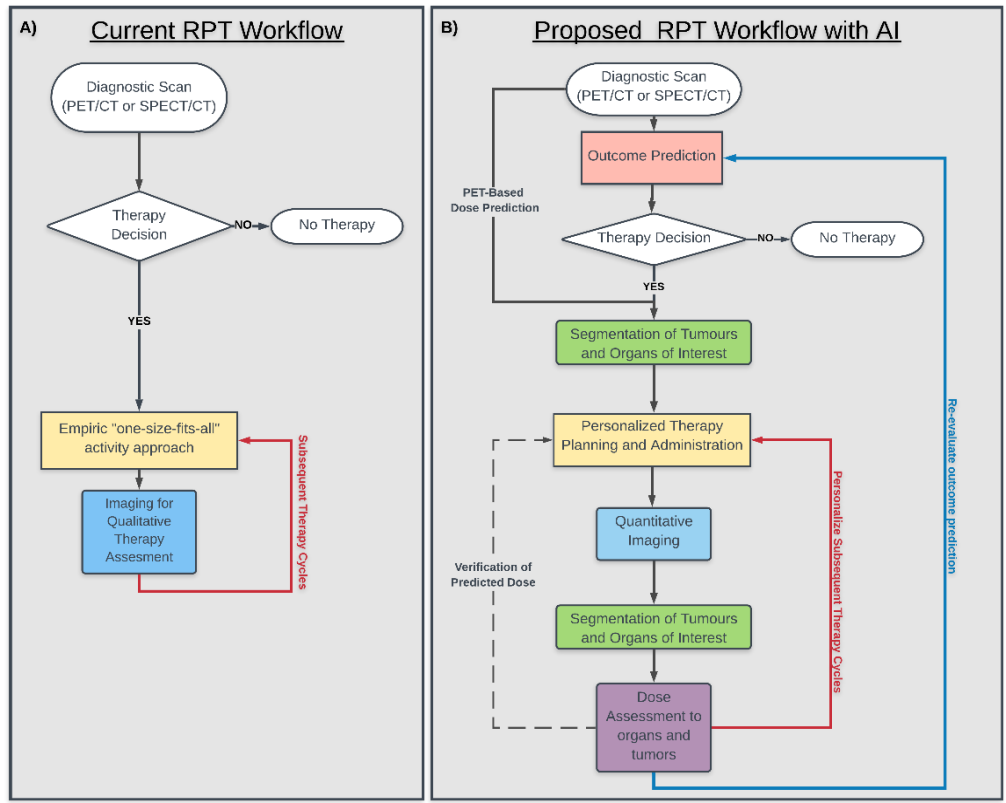

*Figure 2: A) Current typical workflow of RPTs in which dosimetry is not routinely implemented. This only requires a diagnostic examination to establish the suitability of a patient for therapy and possibly qualitative images to determine how good or bad the treatment is performing. B) This diagram represents our vision for the whole theranostics approach. AI is a tool that can assist in every step in this workflow. Even more importantly, AI could predict outcomes and absorbed doses from pre-therapy diagnostic scans to personalize the treatment starting from the first cycle.*

Our vision for a comprehensive theranostics framework (Figure 2B) involves the use of AI to simplify and motivate the personalization of RPTs. AI not only has direct applications in the different steps that form the dosimetry workflow (Figure 3), but could potentially be used to predict outcomes and absorbed doses.

In this work, we aim to highlight areas of importance on which AI can play a very significant role in dosimetry using the theranostics approach. First, we focus on the current challenges of dosimetry after the administration of the therapeutic radiopharmaceutical and discuss the related AI applications. Later, we describe our view on how AI can move us towards personalized RPTs making the theranostics workflow proposed in Figure 2B a reality.

Image-based dosimetry in RPTs

The goal of internal dosimetry is to assess the radiation dose absorbed in healthy and malignant tissue. Report 85 of the International Commission on Radiation Units and Measurements (ICRU) [10] defines absorbed dose caused by the interactions of ionizing radiation in organs and tumors as the amount of energy deposited per unit mass of tissue.

The Committee on Medical Internal Radiation Dose (MIRD) has defined guidelines to estimate absorbed dose in RPT using quantitative images acquired at different time points following administration of a therapeutic radiopharmaceutical [11]. These images are used to measure the radiopharmaceutical biodistribution over time [12]. The workflow for image-based dosimetry includes several processing steps [13] that are illustrated in Figure 3 and correspond to the same colored boxes of Figure 2B. Below, we discuss each of the different steps required for accurate absorbed dose assessments and make recommendations on how AI can further assist throughout the workflow.

Figure 3

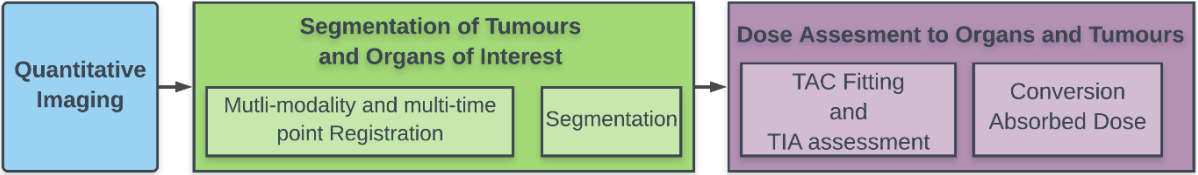

*Figure 3 Schematic representation of the dosimetry workflow for any image-based absorbed dose estimation for radiopharmaceutical therapy.*

## The role of AI in quantitative imaging

The first step in the dosimetry workflow (Figure 3) is the acquisition of quantitative images that allow for the accurate measurement of activity [14]. The goal is to measure the biodistribution of the radiotracer as a function of time. The number of imaging time points that should be acquired is a compromise between optimization of resources, simplification of protocols, and the accuracy for which the patient-specific effective half-life of the radiopharmaceutical can be estimated [15]. Quantitative imaging also implies the use of standardized acquisition protocols, image reconstruction parameters, and methods to determine the camera calibration factor [11, 16].

Both, single photon emission computed tomography (SPECT) and positron emission tomography (PET), are quantitative imaging modalities that allow us to measure radioactivity distribution in the patient over time. Image acquisition and reconstruction parameters needed for accurate quantification is a topic of ongoing research.

Recent work has assessed the reduction in the number of acquired SPECT projections to reduce scan time without compromising quantitative accuracy or image quality. Rydén et al. [17] used a deep convolutional U-net-shaped neural network to generate intermediate $^{177}$Lu SPECT projections (i.e. projections that were not acquired). They found that adding the projections generated by the U-net to the sparsely acquired projections provided similar visual image quality compared to the reference of a full set of projection data. Furthermore, they found comparable kidney activity concentration compared to the one measured from the reconstructed image using a full set of projections. The main advantage of this method is that it allows to scan patients in a much shorter acquisition time. Other investigations have suggested the reduction in acquisition time per projection or the total number of acquired projections in myocardial perfusion SPECT may be compensated for using a deep residual neural network [18].

AI has also been used to generate quantitative images with PET. Studies involving less injected activity or faster acquisitions have been performed [19-21].

Other studies have focused on improvements in image reconstruction [22-26]. Image degrading effects such as scatter and attenuation need to be corrected for to obtain quantitative images. Scatter correction remains a challenging task in SPECT reconstruction, especially for the imaging of pure-beta emitters that do not create any gamma emissions in their decay chain (e.g. Yttrium-90). In these scenarios, the detected energy spectrum of the photons corresponds to the Bremsstrahlung photons. Xiang et al. [27] used Monte Carlo (MC) simulated phantom data to create projections and understand the scatter components. They used this dataset to train a deep convolutional neural network (CNN) that estimated the scatter component in the projections. The CNN estimated scatter was compared to the one derived from MC simulations. The results were very similar between MC and CNN with the advantage that the latter required only a mere fraction of time compared to MC. The use of a fully connected CNN for SPECT reconstruction was investigated by Shao et al. [28] and outperformed conventional ordered subset expectation maximization (OSEM) SPECT reconstruction in terms of image resolution and quantitation.

In PET, new state-of-the-art reconstruction algorithms such as the block sequential regularized expectation maximization algorithm (BSREM) allow for a higher number of iterations without amplifying the image noise [29]. However, the increased number of iterations also increases the time needed to generate an image. AI has been used to speed up the reconstruction by generating images for intermediate iterations [30]. The improvements of reconstruction of newly introduced total body PET images using deep learning (DL) methods is subject to ongoing research [31].

Image denoising allows for the reconstruction of quantitative images with less injected activity, faster acquisition times, or with a higher number of iterations in the reconstruction algorithm. There have been studies showing denoising methods using AI CNNs with scintillation cameras data [32], using generative adversarial networks (GAN) [33] for PET, and using coupled U-Nets for SPECT [34].

The interest in targeted alpha therapies [35] is rapidly increasing, though the quantitative imaging remains a challenge [36]. AI methods could be applied to improve both, image quantification accuracy and quality.

## The role of AI in image registration and segmentation

The positioning of the patient during pre- and post-therapy scans is highly variable. To match the different organs and tumors' radiopharmaceutical uptake between imaging points, accurate image registration is required (Figure 2 and Figure 3). Moreover, anatomical changes (e.g. tumor shrinking or disease progress) between time points requires non-rigid registration methods to fully account for those changes.

### Multi-modality and multi-time point registration

Medical image registration is necessary for subsequent segmentation, treatment planning, image-guided radiotherapy and response assessment [37]. Different imaging modalities such as magnetic resonance (MR), computed tomography (CT), PET and SPECT exhibit differences in resolution and provide different complementary information (i.e. anatomical vs. functional). In addition to discrepancies of patient positioning, intra-abdominal organ movement can occur between the CT and the PET or SPECT acquisition within a single examination. Co-registration of serial and multi-modality images, however, is a challenging task within the dosimetry workflow.

Conventional registration techniques such as rigid, non-rigid [38] and multiresolution approaches [39] can be used for this task. Since image deformation may lead to changes in organ/tumor volumes, that get reflected in mass and activity measurements, it has a direct impact on the estimated absorbed dose. The registration method must be chosen and validated carefully. Significant differences have been shown in absorbed dose estimates, depending on whether manual, rigid or deformable registration methods are applied [38].

AI techniques have shown better accuracy and robustness compared to conventional registration methods and can be generalized better across different modalities [40]. Moreover, AI approaches can mitigate the effects of image artifacts on registration results [41]. Despite the fact that most AI-based registration techniques have not been developed specifically for RPT applications, they have the potential to support the radiation therapy workflow [42].

DL using CNNs has been used for medical image registration using supervised and unsupervised schemes. For instance, supervised training of a convolutional stacked auto-encoder was proposed by Wu et al. [40] to learn discriminative features of images from different modalities. These features were then used in iterative deformable registration.

An unsupervised AI-based registration method was proposed by de Vos et al. [43] and Shan et al. [44] using a CNN without the need to include ground truth labels. Liao et al. [45] proposed a method based on CNNs and reinforcement learning for CT to cone-beam CT registration. Studies on synthetic images (using GANs) with known labels to train a deep registration model without the need of annotated data also exist [46]. A 3D unsupervised network that utilizes a metabolic constraint function (MCF) and a multi-modal similarity measure for PET/CT image registration was proposed by Yu et al. [47] (Figure 4). The MCF is defined based on the standard uptake value (SUV) distribution of hypermetabolic regions to reduce the distortion on the displacement vector field (DVF). The DVF is estimated using a 3D CNN. 3D PET images are then wrapped to 3D CT images by a spatial transformer. The spatial and frequency domain similarity is then calculated based on the registered PET patches and the original CT patches. The loss function of the registration framework is the weighted sum of spatial and frequency similarity and a smoothness of DVF. A similar architecture could potentially be applied to the SPECT/CT data acquired during therapy (Figure 2) and to register diagnostic PET images to the therapeutic images.

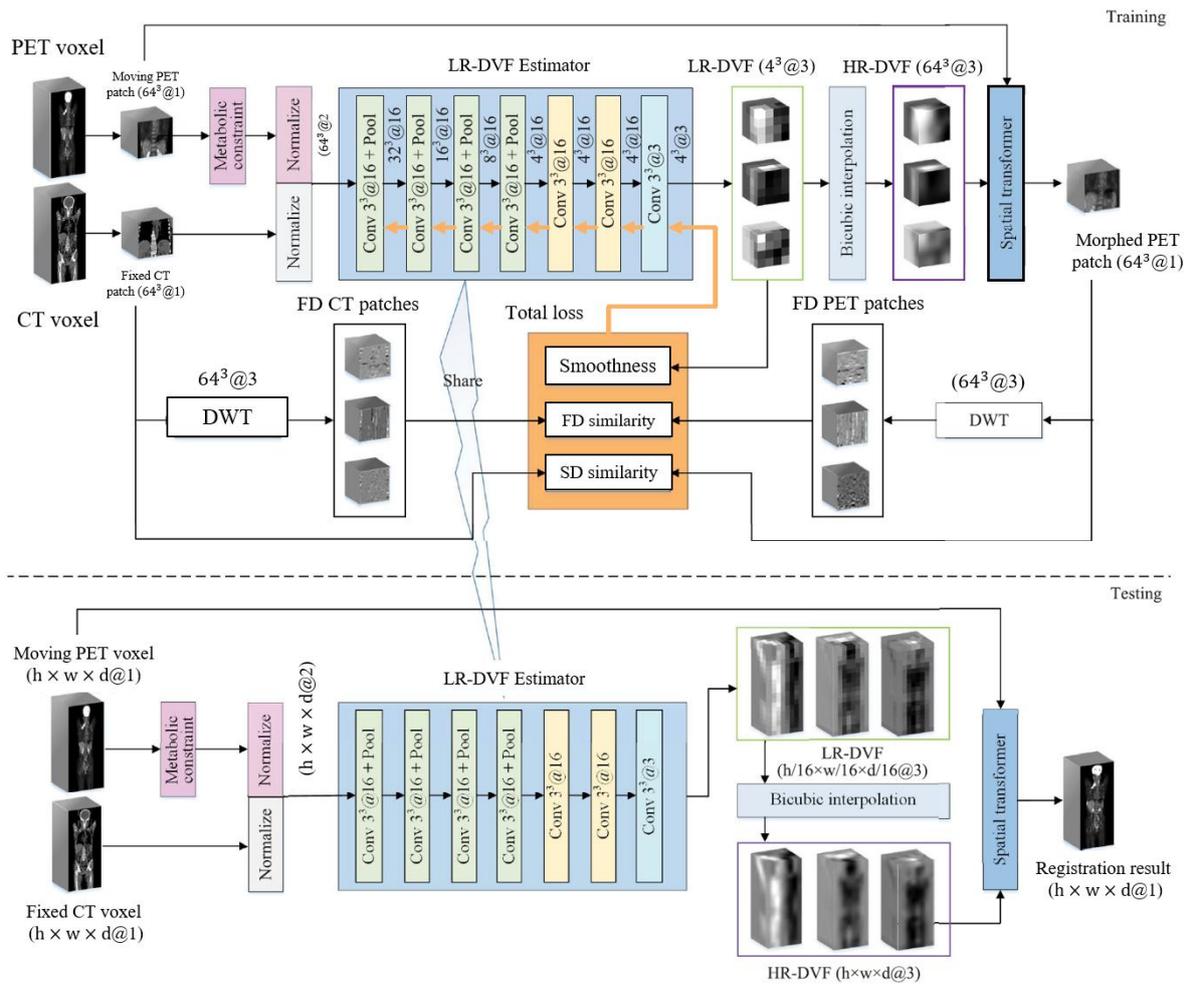

*Figure 4: The unsupervised 3D registration framework proposed by Yu et al. [47]. The figure has already been published under the Creative Commons License which allows us to redistribute it in this document. A copy of the license can be found in https://creativecommons.org/licenses/by/4.0.*

Recent investigations by Guerra et al. [48] for radioembolization purposes used two different CNNs for automatic liver segmentation on MR and CT images respectively, and subsequently registered the segmentation results. We hypothesize that these AI approaches can be used in the future in a RPT context with multiple time point multi-modality images by using CNNs for segmentation and subsequent VOI co-registration.

Segmentation of organs and tumors

The identification of OARs and tumors on images is important for absorbed dose estimation. The segmentation of tumors is required for tumor dosimetry that is a critical component in determining the treatment response of RPT [49]. However, segmentation is the most time-consuming task in the dosimetry workflow (Figure 2 and Figure 3) since it often relies on manual delineation of volumes of interest (VOI) [50-52]. Segmentation allows the measurement of activity within each organ and tumor as well as the estimate of the corresponding mass of each VOI. Both quantities are required for accurate dosimetry calculations.

Compared to external beam radiotherapy, it is even more challenging for manual segmentation of tumor lesions for RPT, which specifically treats patients with metastatic cancer. Often patients may have a large number of lesions across the body, of heterogeneous sizes and tracer uptakes. Segmentation of all these lesions manually is not practical. Manual segmentation is also subject to intra- [53] and inter-observer variability [54]. Validated AI-based models for fully automated, robust, accurate segmentation of organs/lesions in PET, PET/CT and SPECT/CT images can help delineate OARs and lesions to achieve a personalized dosimetry framework. Normal organ segmentation approaches using DL models could use CT data [55, 56], or combined image data such as PET/CT [57, 58]. Wang et al. [59] segmented normal organs based on CT images using a multi-atlas method and refined the segmentation on the PET images. A triple-combining 2.5D U-Net, which simultaneously extracts features from axial, coronal and sagittal planes, has been developed to mimic the workflow of physicians for the automated characterization of lesions on PSMA PET [60].

Diagnostic PET images can be expected to have similar intensity profiles as SPECT images acquired in therapy because they are targeting the same receptors. This allows using transfer-learning approaches for segmentation. Diagnostic PET images should be smoothed in this regard to account for the differences in resolution with respect to SPECT. The quality of CT images is the same or very similar between PET/CT and SPECT/CT modalities. AI models can be pre-trained on PET/CT images and then "tuned" using SPECT/CT data. The cross-modality knowledge transfer for lesion segmentation in SPECT images using PET segmentations can be done using unsupervised adversarial training to learn feature mapping between domains (PET and SPECT) as previously shown for domain adaptation from MR to CT [61]. The Probability map (PM) based on diagnostic PET images can be estimated and added to the segmentation model for SPECT/CT images. This learnt PM captures the probability that a voxel in a SPECT image belongs to a tumor or OAR. The PM may not be accurate enough to fully segment SPECT/CT images (as tumors can change over time) but can be used as initial guidance (e.g. increase probability of detection for smaller tumors).

In the direct context of segmentation for dosimetry, Jackson et al. [62] showed promising results when using a 3D CNN for kidney segmentation on the low-dose CT from post-therapy [$^{177}$Lu]Lu-PSMA SPECT/CT against manual organ delineation for renal absorbed dose estimation. In addition, a 3D U-net model was proposed for kidney segmentation for uptake quantification [63]. Tang et al. [64] suggested a CNN model for liver segmentation be used for personalized liver radioembolization.

Besides intra-therapy cycle image registration and segmentation, the possibility to transfer VOIs to subsequent cycles should be investigated. AI could further assist in registering and segmenting intra-abdominal organ movement and tumor shrinkage or disease progress between therapy cycles.

## The role of AI in time activity curve assessment and time-integration of activity

Following registration and segmentation, the next step of the dosimetry workflow pertains to the fit of a model function to the time activity curve (TAC) (Figure 3) on an organ or voxel level. This model function must be chosen carefully to describe the pharmacokinetics of the radiopharmaceutical under investigation. Typically, mono-exponential or bi-exponential functions are used (Figure 5) but tri-exponential functions have also been found in literature [65]. Also, there are situations in which the initial uptake is approximated using a trapezoid. The subsequent calculation of the time-integrated activity (TIA) can be performed on an organ or voxel level. The latter yields a 3D time-integrated activity map (TIAM) [66], from which a 3D absorbed dose estimation can be completed.

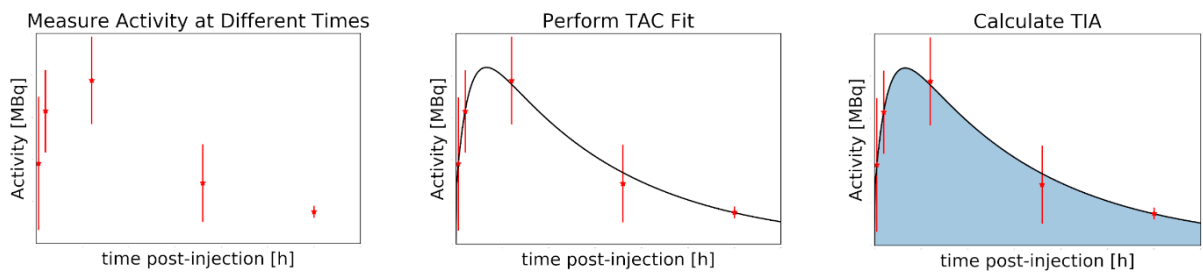

*Figure 5 Simplified representation of the derivation of TIA: Measurement of activity over time at discrete points; fit of mono-exponential decay function to data points; calculation of the area under the curve, i.e. time-integrated activity.*

The fit function describes the pharmacokinetics of the radiopharmaceutical. Sarrut et al. [66] described a multimodal fitting approach on the voxel level for multiple fitting models using nonlinear least square optimization. The best fitting model per voxel was then chosen based on the Akaike information criterion similar to the proposed one in the NUKFIT software by Kletting et al. [67]. The applicability of a particle filter to denoise TACs on the voxel level was proposed by Götz et al. [68] with promising results. Kost et al. [69] on the contrary used a different approach of first generating absorbed dose rate maps on each of the serial quantitative activity images at the voxel level followed by pharmacokinetic modelling.

AI remains to be actively used for TAC or TIA estimation with great potential in this step of the dosimetry workflow both in organ or voxel level approaches. Possible applications could include investigation of CNNs that use information of serial quantitative images to predict TIAs using only a single post-therapy scan. Moreover, data from the diagnostic imaging could be used in conjunction with the single therapeutic image acquisition to improve the TAC and thus have higher confidence in the absorbed dose results further down the workflow. Lastly, AI can use the information from the diagnostic scans and therapeutic cycles together to improve the TIA of subsequent therapy cycles. Reducing the number of post-therapy scans is advantageous for the patients' comfort and decreases the workload for clinical personnel. Advances in image co-registration and reconstruction may further enhance progress in minimizing error from voxel-wise fitting due to image artifacts in individual voxels.

## The role of AI in conversion to absorbed dose

The last step of the workflow entails the conversion of TIA into absorbed dose (Figure 3). Dosimetry can be performed on an organ [70] or voxel level [71, 72], each associated with different degrees of accuracy and complexity.

Organ level absorbed dose estimation according to MIRD [70] uses organ- and radionuclide specific S-values derived from simulations with reference human phantoms. These S-values yield the absorbed dose in a target organ per decay in a source organ. The radiation absorbed dose to an organ is hence derived from the sum of all sources to target combinations of TIAs multiplied by the respective S-values.

Similarly, voxel S-value kernels are simulated for a specific tissue composition and radionuclide. Kernels are then convolved with the TIAM to create 3D absorbed dose maps [71]. Monte Carlo (MC) simulations use the patient-individual 3D CT and TIAM to precisely model the absorbed dose for heterogeneous tissues and activity distributions [72]. Whilst MC simulation-based dosimetry taking into account heterogeneous activity and tissue distributions is still the gold standard against to which other methods are validated [73, 74], this is the most complex, computationally demanding and time-consuming dosimetry method. AI offers the potential to maintain accuracy of MC dosimetry while reducing the time required. Table 1 summarizes the different assumptions made when different dosimetry approaches are used.

| Type of \ When using | Organ S-value | Voxel S-value | MC dosimetry simulation |
|---|---|---|---|
| Activity distribution | Homogeneous | Heterogeneous | Heterogeneous |
| Tissue composition | Homogeneous medium | Homogeneous medium | Heterogeneous patient anatomy using CT |
| Dose Output | Mean absorbed dose | 3D absorbed dose map | 3D absorbed dose map |

Table 1: Different assumptions per dosimetry method and yielded absorbed dose result (i.e. organ or voxel level). The complexity as well as the accuracy of dosimetry methods increases from left to right.

Image-based organ or voxel level dosimetry approaches yield macroscopic absorbed doses. However, there is increasing interest in describing the radiation damage on smaller region of an organ or tumor or even at the cellular level [75]. Knowing this can provide a better understanding of the underlying radiobiological effects during RPT [76-78]. Currently, RPTs are limited to absorbed doses based on the experience of external beam radiation therapies (EBRT). However, differences in absorbed dose rates and number of cycles of RPT compared to EBRT has lead us to think that different absorbed dose limits should be set for internal radiation therapies. As an example, the absorbed dose to kidneys is commonly limited to 23 Gy. However, absorbed doses of up to 40 Gy have been shown to be tolerated by patients without risk factors [79]. AI can potentially be used to combine multi-scale dosimetry knowledge for accurate effective dose modeling. AI may unveil the complex relationship between pre-therapy patient data, such as imaging, demographic data, lab results, and the radiation dose distribution to be obtained during therapy, which is a problem too complicated to be described by conventional mathematical modelling approaches. GANs attempt to model the post-therapy voxel-wise dosimetry directly from pre-therapy imaging.

The prediction of deposited energy distribution and voxel-based dosimetry using a deep neural network was assessed by Akhavanallaf et al. [51] as illustrated in Figure 6. Their approach used whole-body 3D density maps (derived from patient CT images) and 3D absorbed dose maps (generated with MC simulations) as input to train a deep neural network to generate tissue-specific S-value kernels. Their method has the potential to overcome the general limitation of voxel S-value kernels that assume homogenous tissue and typically water density. Götz et al. [80] used a CNN to predict density specific voxel S-value kernels.

As alternative to MC dosimetry, Lee et al. [81] studied the use of a CNN for dosimetry estimation at a voxel level. Their network was trained to yield absorbed dose rate maps for activity and tissue distributions of Gallium-68 ($^{68}$Ga) [$^{68}$Ga]Ga-NOTARGD PET/CT based on ground truth MC simulation derived absorbed dose rate maps. The dose difference of CNN derived absorbed dose rate maps against MC was below 2% with a time-effort of less than 4 minutes compared to over 235 hours computation time of the MC simulation. Similarly, Götz et al. [82] trained a CNN with MC reference absorbed dose maps to generate voxel absorbed dose maps on the input of density maps from CT and TIAMs from serial $^{177}$Lu SPECT/CT. The predicted absorbed dose maps from the model outperformed the use of a soft tissue voxel S-value kernel when compared to MC generated absorbed dose maps.

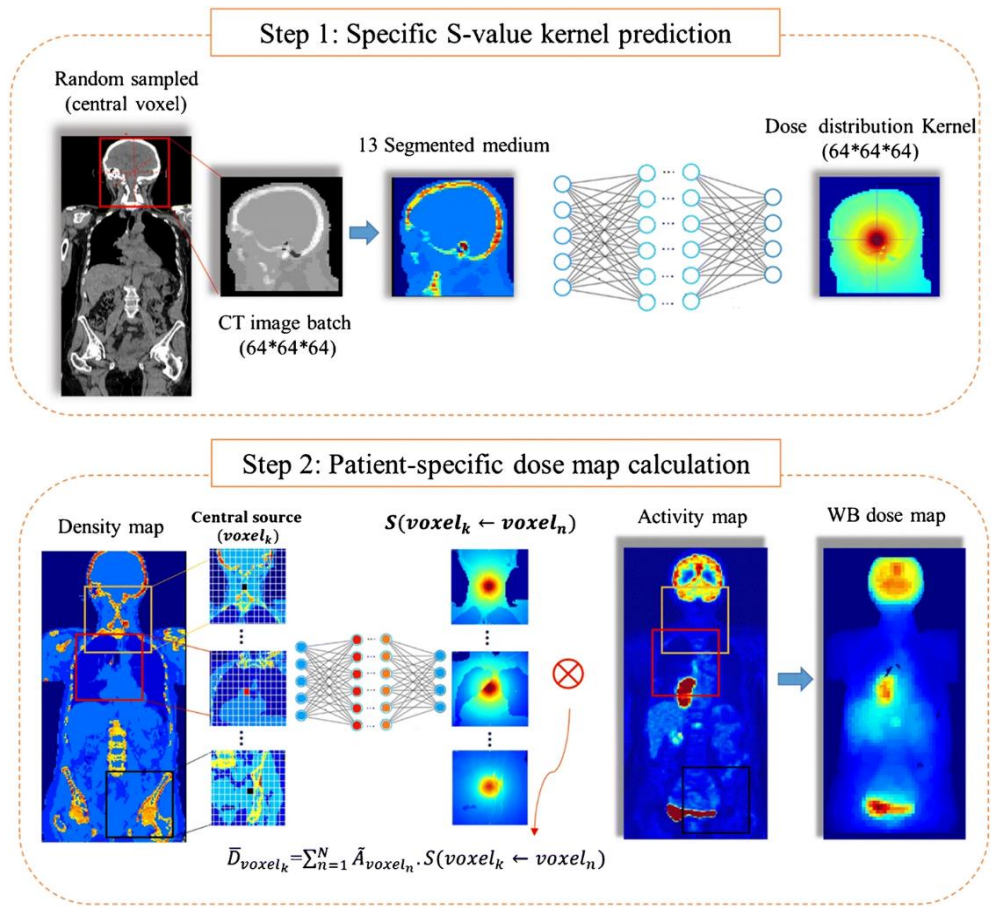

*Figure 6: Diagram of the procedure used by Akhavanallaf et. al. [51] in which a dose kernel is generated using a deep neural network. The kernel can then be convolved with the TIAM to generate an absorbed dose distribution. The figure has been already published under the Creative Commons Attribution 4.0 that allows us to redistribute it in this document. A copy of the license can be found in https://creativecommons.org/licenses/by/4.0.*

## AI and the future of personalized RPT: radiomics, dosiomics and outcome prediction

Improved therapeutic absorbed dose estimation directly translates into the possibility of correlating absorbed dose with tumor response or normal organ toxicities. Knowledge of absorbed dose-response relationships might enable us to personalize activity planning for subsequent therapy cycles. Combined analysis of diagnostic imaging and therapeutic radiation absorbed doses might then allow for therapy outcome prediction.

Because of the theranostics approach, the outcome prediction could potentially be implemented with the diagnostic scans even before the RPT. This prediction can then be verified and updated in combination with additional information collected in subsequent therapy cycles (Figure 2B). Studies like the one performed by Xue et al. [83] have already used voxel-wise absorbed dose prediction for [$^{177}$Lu]Lu-PSMA therapy based on pre-therapeutic PSMA PET/CT. They trained GANs using the diagnostic [$^{68}$Ga]Ga-PSMA PET/CTs and 3D absorbed dose maps. This described approach in combination with known dose-response relationships could assist the physicians in making the best therapy decision (Figure 2B). AI has shown the capability to discover effective predictive biomarkers for treatment outcome and long-term survival. There is an untapped potential to apply radiomics analysis to molecular imaging (both from pre- and post-therapy images) that can contain biological information. Moreover, there might be features detectable from within the 3D absorbed dose maps that can also show value towards better understanding of therapy response and outcome, enabling further personalization of therapies. We refer to the analysis performed on the 3D absorbed dose maps using the term dosiomics. The combination of radiomic features from CT, PET, MRI, and SPECT with dosiomic features from the absorbed dose maps can be used to train AI models that can better guide physicians with treatment planning and absorbed dose predictions. Moreover, deep learning approaches using different modalities of imaging and absorbed dose maps can inherently find features that are good predictors of outcome.

To develop robust outcome prediction models based on radiomics and dosiomics, the datasets must be representative of the disease and contain variant types and severities of it. Standardized imaging protocols and pre-processing steps are important to ensure consistent image quality [84]. An array of features can be used as input to the models for outcome prediction. For example, (i) absorbed dose-volume histogram measures could be computed from segmented 3D absorbed dose maps and are increasingly available in radiomics software packages. Such measures could also be correlated with tumor control probability and potential normal tissue toxicity. (ii) Quantitative features from diagnostic scans such as $SUV_{mean}$, $SUV_{max}$, $SUV_{peak}$, molecular tumor volume, total lesion activity (TLA), and total lesion fraction (i.e. TLA divided by body weight) could further serve as input for outcome prediction models. Analysis can be performed using PET-only, SPECT-only, and PET/SPECT. The subtraction of these parameters between cycles could be applied for outcome prediction [85, 86]. (iii) The analysis can include an array of radiomics features, beyond the above-mentioned simpler metrics. It has been shown that radiomic features at PET resolution can preserve their value at lower SPECT resolution [87].

Investigations can include the detection of radiomic features in relation to biomarkers for disease staging [88] and could be extended to use of neural networks (NNs) or CNNs to predict the outcome of therapy [89-91]. In addition, NNs can model nonlinear survival data by classifications [92]. A deep network can directly extract and identify the most predictive radiomic features and could further learn unique features that may not be captured by handcrafted radiomics. The new paradigms of fusion radiomics, have been investigated on PET/CT in head & neck cancer [93, 94] and could be extended to SPECT/CT scans. Furthermore, subsequent therapy cycles could involve adapted therapy planning based on available dosimetry and outcome modeling.

As technology keeps improving, it also expands the possibilities of data collection. For example, recently developed total body PET scanners [95, 96] would enable the collection of dynamic whole-body data [97, 98] of the diagnostic radiopharmaceutical. This allows for the generation of parametric images that also provide information about the biokinetics of the radiopharmaceutical [99]. Although there are limitations related to the early acquisition time and shorter half-life of diagnostic radionuclides compared to therapeutic radionuclides, AI can play an important role in understanding the dynamic scans and can possibly predict the uptake of the radiopharmaceutical in the therapy cycle.

In addition, the denoising examples mentioned before, in combination with more sensitive scanners, can allow us to perform the diagnostic scan at much later times that might correspond to the washout phase of the radiotracer. For example, it has been reported that the new EXPLORER total body PET scanner has the ability to image a patient injected with [$^{18}$F]FDG up to 5 half-lives after injection [96]; something unthinkable with current limited axial field of view scanners. Also, longer half-life PET radionuclides such as Copper-64 ($^{64}$Cu) (12.7 h half-life) or Zirconium-89 ($^{89}$Zr) (78.4 h half-life) that are used to label theranostic pairs, could provide the data required to predict absorbed doses and outcomes for which AI is a fantastic tool to explore with currently existing scanners.

Benefiting of the emerging research and applications of AI in the fields of quantitative imaging, segmentation, registration absorbed dose prediction, and outcome modelling we believe personalized therapies can easily be implemented in the clinical setting.

## Conclusion

For adaptive RPT planning and personalized activity prescription, predictive dosimetry prior to treatment as well as absorbed dose verification is required to optimize therapy. For that, it is mandatory to have standardized protocols and reliable absorbed dose values first. Hence, efforts should concentrate on accuracy improvements of any of the steps within the dosimetry workflow. Cancer treatments are often difficult and complex, but the nuclear medicine community can incorporate the technological advancements of AI to make dosimetry a feasible task in the clinical setting. This includes applications for image quantification, registration, segmentation, biodistribution modeling, and absorbed dose value calculation. Predictive modelling of therapy outcome and absorbed doses following a therapeutic injection can assist in treatment planning and benefit patients from personalized RPTs.

The future of personalized radiopharmaceutical therapy will likely benefit from active utilization of AI methods in the field of theranostics. This work highlighted different possible applications of AI, with the hope to motivate the community to expand and align efforts towards routine and reliable personalization of RPTs.

**Acknowledgements:** This work was in part supported by the Natural Sciences and Engineering Research Council of Canada (NSERC) Discovery Grants RGPIN-2019-06467 and RGPIN-2021-02965.

## List of Abbreviations

| | |
|---|---|
| AI: | Artificial intelligence |
| BSREM: | Block sequential regularized expectation maximization algorithm |
| CNN: | Convolutional neural network |
| CT: | Computed tomography |
| $^{64}$Cu: | Copper-64 |
| DL: | Deep learning |
| DVF: | Displacement vector field |
| EANM: | European association of nuclear medicine |
| EBRT: | External beam radiation therapy |
| $^{18}$F: | Fluorine-18 |
| GAN: | Generative adversarial network |
| IAEA: | International atomic energy agency |
| ICRU: | International commission on radiation units and measurements |
| $^{177}$Lu: | Lutetium-177 |
| MC: | Monte carlo |
| MCF: | Metabolic constraint function |
| mCRPC: | Metastatic, castration-resistant prostate cancer |
| MIRD: | Medical internal radiation dose |
| MR: | Magnetic resonance |
| NET: | Neuroendocrine tumor |
| NN: | Neural network |
| OAR: | Organ at risk |
| OSEM: | Ordered subset expectation maximization |
| PET: | Positron emission tomography |
| PM: | Probability map |
| PSMA: | Prostate-specific membrane antigen |
| RPT: | Radiopharmaceutical therapy |
| SNMMI: | Society of nuclear medicine and molecular imaging |
| SPECT: | Single photon emission computed tomography |
| SUV: | Standard uptake value |
| TAC: | Time activity curve |
| TIA: | Time-integrated activity |
| TIAM: | Time-integrated activity map |
| TLA: | Total lesion activity |
| VOI: | Volume of interest |